\documentclass[preprint,showpacs,preprintnumbers,amsmath,amssymb]{revtex4}

\usepackage{graphicx}
\usepackage{dcolumn}
\usepackage{bm}
\begin{document}

\def\Sgr{{\bf S} }
\def\Tgr{{\bf T} }
\def\Zgr{{\bf Z} }
\def\sgr{{\bf s} }
\def\sigr{{\bf \sigma} }
\def\1gr{{\bf 1} }
\def\brak{\langle}
\def\kket{\rangle}
\def\be{\begin{equation}}
\def\ee{\end{equation}}
\def\al{\alpha}
\def\bet{\beta}
\def\ga{\gamma}
\def\del{\delta}
\def\Tr{{\rm Tr\,} }
\def\dg{\dagger}

\author{Jacek Wojtkiewicz}
\affiliation{Department for Mathematical Methods in
Physics\\Ho\.za 74, 00-682 Warsaw, Poland\\e-mail: wjacek@fuw.edu.pl}
\title{
 SOME PROPERTIES OF FRUSTRATED SPIN SYSTEMS:  EXTENSIONS AND
 APPLICATIONS  OF LIEB-SCHUPP APPROACH 
}

\begin{abstract}
Lieb and Schupp have obtained, using certain version of
``spin-reflection positivity'' method,  a number of ground-state
properties for  
frustrated Heisenberg models. One group of these results is related to
singlet nature of ground state and it needs an assumption of
reflection symmetry present in the system. In this paper, it is shown
that the result holds also for other symmetries (inversion etc.). The
second Lieb-Schupp result is relation between ground-state energies of
certain systems. In the paper, this relation is applied to
multidimensional models on various lattices. 
\end{abstract}
\maketitle

\section{Introduction}
\begin{figure}
\includegraphics{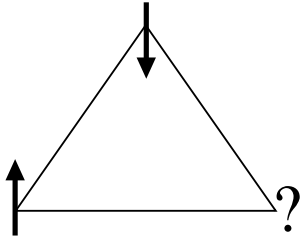}
\caption{\label{fig:fig1}  Example of frustrated spin system. }
\label{fig1}
\end{figure}

Geometric frustration takes place, when no arrangement of spins on the
lattice is possible in such a way that all interactions minimize their
energy (see FIG.~\ref{fig1}). The canonical example is an antiferromagnetic
Ising model on 
triangular lattice \cite{WannHout}. Another systems where the
geometric frustration is particularly strong, are antiferromagnetic
systems on kagom\'{e}, pyrochlore, or (in d=3) fcc cubic lattices.  

Frustrated systems are both very interesting and hard to analyse and
understand, both in classical version and especially in the quantum
case. The source of these difficulties traces back to the large
ground-state degeneracy in the classical version. Such systems are
very sensitive to perturbations. A consequence  is possibility of
very complicated phase diagram at finite temperatures. This situation
takes place, for
instance, in an ANNNI model \cite{FishSzp} (infinite number of phases,
devil's staircase, etc). Besides of numerous efforts
and important results, \cite{FishSel}, \cite{FishSzp}, \cite{DinMaz}, 
 (for reviews, see \cite{Selke}, \cite{FrIs}, \cite{FrSS}),
 full treatment of such systems is not worked out so
  far. 

The situation for quantum frustrated antiferromagnets is even less
clear. It is generally suspected that the ground state emerging as a
linear combination of many classical configurations is a
featureless, ``spin liquid'' state, i.e. state without
  long-range ordering, where 
 correlation  functions fall-off exponentially. However, one cannot
 exclude another scenario: ``order by disorder'' -- exotic orderings
 absent in a classical version of these models. Such scenarios are
 moreover sensitive to the underlying lattice structure. (For a
 review, see for instance
 \cite{Moessner}). To my best knowledge, no general definite
 conclusions have been obtained here. 

Most informations obtained comes from analysis using approximate
methods, such as guessing of ground-state wave functions, numerical
diagonalization of small systems, spin waves, semiclassical
(i.e. large-S) approximations.
However, it is difficult to estimate how reliable are these
approximations. For this reason, exact results are very
desirable. However, they are very exceptional.

In such a situation, exact results obtained by Lieb and Schupp
\cite{LS1}, \cite{LS2} (summarized also in \cite{S3})
for fully frustrate systems  are of first
importance. They are  interesting both as themselves
and moreover, they can  serve as a test of validity of  approximate
methods.   Results obtained in
\cite{LS1}, \cite{LS2} can be divided into two groups. The first one
concern ground-state properties for such systems and it
can be summarized as follows: ground states are
singlets; zero-field magnetization is zero; susceptibility is bounded
by certain constant. One of assumptions of these theorems is {\em
  reflection symmetry}. It was apparent that this method could work also
  for another kind of symmetry, not necessarily reflection one. Such a
generalization turned out to be possible, and this is the first result
of the paper.

The second collection of results \cite{LS2} concern comparison of
ground state energies for spin rings. It turned out that similar
results hold for more complicated systems (multidimensional
lattices). This is second result of the paper.

The outlook of the paper is as follows. In Sec.~\ref{SecLSandGen}, the basic
ingredients of Lieb and Schupp technique is presented. Some systems
not mentioned in \cite{LS1}, \cite{LS2}, for which they results hold,
are listed. Then modifications of their technique to systems
exhibiting symmetry other than reflection one (rotation or inversion)
are described. In Sec.\ref{SecCompar}, Lieb and Schupp
results concerning comparison of ground-state energies for spin rings
with the use of certain matrix inequality are described. Then it is
discussed  how
this technique can be applied to other systems -- for instance
g.s. energies of systems on different  lattices. Section
Sec.~\ref{SecSumm}  contains short summary and
conclusions. The Appendix contains proof of Lieb-Schupp inequality,
which is fundamental for Sec.~\ref{SecCompar}.


\section{Lieb-Schupp approach and generalizations for spin systems
 with symmetries}
\label{SecLSandGen}
\subsection{Reflection-symmetric systems}
\label{Subsec:2.1}
\subsubsection{Assumptions}
\label{Subsubsec:2.1.1}
We make the following assumptions concerning systems under consideration.
\begin{enumerate}
\item We consider Heisenberg models for arbitrary spin. (Below we consider
mainly the  $s=1\slash 2$ case, but generalization to other spin values 
is straightforward).
\item The system is invariant with respect to the reflection about the
 $O$ axis 
 ($d-1$ plane for $d$-dimensional system). The system consist of two parts
("Left" and "Right" ones), which are interchanged under reflection
 (see FIG.~\ref{fig2}). The 
Hamiltonian is a sum of three parts:
 $H_L, H_R$ and $ H_C$;  $H_L$ acts only on the left part
 (i.e. $H_L=h\otimes \1gr$),
 whereas 
 $H_R$ acts on the right one (i.e. $H_R = \1gr\otimes h$).  
$H_C$ is the Hamiltonian for "crossing bonds" (i.e.
for bonds which intersect the symmetry line). It has the form:
\be
H_C=\sum_{i\in L,\: i'\in R}j_{i,i'}\sgr_i \cdot \sgr_{i'}
\label{PostacHC}
\ee
where $j_{i,i'}$ is a diagonal matrix with non-negative diagonal
elements.  $\sgr_i$ can have various nature; it can be a single 
spin or, more generally, it can have the form: $\sgr_i = \sum_{\al\in L} 
j_{i;\al} \sgr_\al$, where $j_{i;\al}$ are real coefficients. The
objects  $\sgr_i$ can be
different for different indices $i$.
Under reflection operation, $H_L$ is transformed to $H_R$  and vice versa, 
whereas $H_C$ transforms into itself.
\begin{figure}
\includegraphics{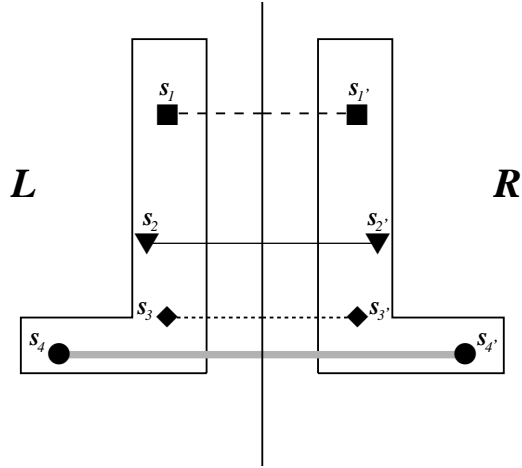}
\caption{\label{fig:fig2}  System exhibiting reflection symmetry.}
\label{fig2}
\end{figure}
\item $H_L$ and $H_R$ are (almost) arbitrary: the only limitations are: both
commute with the total spin operator, and
\item  moreover, operators are {\em real} in the $S^3$ basis.
\end{enumerate}
\subsubsection{Ground state properties}
\label{Subsubsec:2.1.2}
Every state of the system can be expressed as a linear combination:
\be
\psi=\sum_{\alpha,\beta}c_{\alpha \beta} 
\psi^L_\alpha\otimes(\psi^R_\beta)_{\rm rot},
\label{ReprOfState}
\ee
where states $\psi^L_\alpha$ form a real orthonormal base of $S^3$
eigenstates for the left subsystem, whereas  
 $(\psi^R_\beta)_{\rm rot}$ are corresponding {\em rotated} states for
the  right subsystem. (The rotation --  meant as an operation in the
spin space  -- acting for the state: $ |s,m\kket $ produces the state
$ (-1)^{s-m}|s,-m\kket$).

The eigenvalue problem: 
$H\psi=E\psi$  can be written as a matrix equation for $c$ matrix: 
\be
h_L c+c(h_R)^T -\sum_{i=1}^3\sum_{y} t_y^{(i)} c (t_y^{(i)})^T=Ec,
\label{EqMac}
\ee
where $(h_L)_{\alpha\beta}$ and  $(h_R)_{\alpha\beta}$ are real, symmetric 
matrices for corresponding terms of the Hamiltonian,
whereas $t^{(i)}_y$ are {\em
  real} matrices, defined for spin operator $\sgr_y = \sum_{\al\in L}
j_{y;\al} \sgr_\al$ 
(in the  $y$-th "bond") by: $t_{y;\alpha\beta}^{(1,3)} = 
\langle \psi_\alpha^L | s_y^{(1,3)}  | \psi_\beta^L\rangle $ and  
$t_{y;\alpha\beta}^{(2)} = 
i \langle \psi_\alpha^L | s_y^{(2)}  |
\psi_\beta^L\rangle$. Notice the total "minus" sign before third term in 
(\ref{EqMac});
it is so because replacement of   $\sgr_y$ by $\sgr_{y'}$
gives sign change for components $1$ and $3$, whereas the 
 $i$ factor in definition of
$t^{(2)}$ gives minus sign for component $2$.

The energy expectation value, expressed in term of the $c$ matrix, is
\be
\brak \psi |H | \psi \kket = 
\Tr (cc^\dagger h_L) + \Tr (c^\dagger c h_R)- 
 \sum_{y,i} \Tr c^\dagger t_y^{(i)} c (t_y^{(i)})^\dagger.
\label{psiHpsi}
\ee
Lieb and Schupp have established  a number of results concerning
ground states of reflection-symmetric systems. There are:
\begin{enumerate}
\item The $c$ matrix is hermitian, i.e. 
 without loss of generality we can express 
eigenstates of $H$  by {\em hermitian} matrix $c=c^\dagger$. (It is proved
using the left-right symmetry and a form (\ref{psiHpsi}) for energy).
\item 
Let the  $c$ matrix corresponds to the ground state; then, also the
$|c|$  matrix corresponds to the ground state. Here $|c|=\sqrt{c^2}$;
we take the unique positive
square root. (It is proved by writing down (\ref{psiHpsi}) in the diagonal
basis).
\item This implies that 
we can choose the base for ground
states as a collection of functions built up from
positive semi-definite (p.s.d.) coefficient matrices.
\item The overlap of the ground state with canonical
spin-zero state (given by unit matrix in a basis of $S^{(3)}$ eigenstates
in either subsystem) is non-zero. 
\item It implies that there exist ground state with spin zero, and
moreover, that coefficient matrix for this state is positive
semidefinite. It follows from the inequality: $E_b\geq E_0$ ($E_b$ is
an energy  of the system in the magnetic field $b$), which in turn is
proved  using the matrix inequality (\ref{MatrIneq}).
\item We have "ice rule" for frustrated units, i.e. for collection of spins
appearing in the expression $\sgr_i \cdot \sgr_{i'}$. More precisely,
if we  denote:
$\sgr_i =\sum_{\al\in L}
j_\al \sgr_\al$, then the expectation of the third component of sites in each
crossing bond vanishes for any ground state $|\psi_0\rangle$:
\[
\langle\psi_0 | \sgr_i + \sgr_{i'}|\psi_0\rangle
\] 
\end{enumerate}
Under additional assumptions, these results can be made stronger. For instance,
if the system has periodic boundary condition in at least one direction, then
{\em all} ground states are singlets.
Moreover, the magnetic susceptibility is bounded from above, both in
the  ground state and in positive temperatures.

{\em Remark.} Lieb and Schupp in \cite{LS1} and \cite{LS2}
have developed general method and (almost)
haven't given examples of systems where their method could apply; the
only example discussed is the
 checkerboard pyrochlore lattice. Some further examples of physical
interest, where their results are applicable, are:
\begin{itemize}
\item $J_1-J_2$ 2d Heisenberg model:
\[
H=J_1\sum_{n.n.} \sgr_i\cdot \sgr_j 
+J_2\sum_{n.n.n.} \sgr_i\cdot \sgr_j 
\]
considered by numerous authors (see, for instance, \cite{J1J2Heis} and 
references therein). This model is very interesting, as when
$J_2\slash J_1$ is varied,  it exhibits 
a transition from ordered, antiferromagnetic state to the 'glassy', 
non-magnetic, spin-liquid-like one. 
Lieb and Schupp results apply for $J_2\leq\frac{1}{2} J_1$. Most of approximate
methods {\em assume} that ground state is singlet. Lieb and Schupp' results
can be used to justify
this assumption. Moreover, it can serve also as a test of these
 methods by supplying
 rigorous upper bounds
for susceptibility.  
\item
Axial Next-Nearest-Neighbour Heisenberg (ANNNH) model \cite{Selke}. This
is Heisenberg model with two coupling constants; we have isotropic
coupling  $J_1$ between nearest neighbours, and moreover,
there is coupling $J_2$ between second neighbours along one of the axes (say,
$z$ axis).
Such models have been used to describe helical and incommensurate
configurations,  and Lifshitz points in magnets (\cite{Selke} and
references therein).
 Lieb-Schupp results can be applied when both constants are
 antiferromagnetic,  
or when $J_1$ is AF and $J_2$ is F (in this case, one should take the
 reflection plane to be parallel to the line formed by $J_2$ couplings).
\end{itemize}
\subsection{Generalization to other symmetries}
\label{Subsec:2.2}
In the course of proofs in the previous Subsection, geometric
properties of systems exhibiting reflection symmetry were {\em not}
employed. Only assumption which was used was that $h_L$ transforms
into $h_R$ and vice versa; particular nature of this transformation
was not essential. It suggests that more general symmetry operations
than reflection are allowed. It is the case; more precise formulation
is as follows.

We make  assumptions identical as 1., 3. and 4. in
Subsec. \ref{Subsubsec:2.1.1} ;
the assumption 2. is changed into the following one.

\begin{figure}
\includegraphics{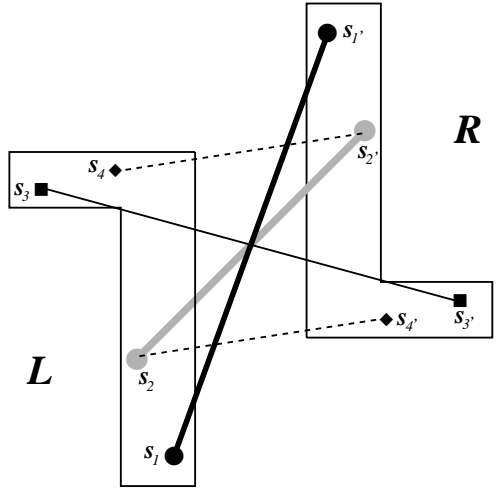}
\caption{\label{fig:fig3}  System exhibiting inversion symmetry.}
\label{fig3}
\end{figure}

 2'. The system again can be divided by two identical parts "L" and "R",
each of them is described by the Hamiltonians $H_L= h\otimes \1gr$ and 
$H_L=  \1gr \otimes h$, respectively. $H_C$ is the 
Hamiltonian for "bonds" between spins in  $L$ and $R$ subsystems.  It
has  the form:
\[
H_C=\sum_{i\in L,i'\in R}\ga_{i,i'} \sgr_i \cdot \sgr_{i'}
\]
where the symmetric $\ga_{i,i'}$ matrix is positive definite.  
One can view on this property as a demand that $H_C$ has to be
``globally antiferromagnetic'', i.e. some coupling constants
$\ga_{i,i'}$ can be negative (ferromagnetic), however, the whole
matrix $\ga$ has to be positive definite.
The whole system is invariant with respect to the some symmetry
operation $\Tgr$,  
such that $\Tgr^2={\rm Id}$; $\Tgr$ transforms the "L" subsystem into
"R" one and vice versa. The following operations: reflection, 
$C_2$ rotation (see FIG.~\ref{fig3}),
or inversion can serve as examples of such operation. Under action of $T$,
$H_L$ transforms into $H_R$, $H_R$ into $H_L$ and $H_C$ into itself.

The system described above can be easily transformed into form equivalent
to the one considered in the previous Subsection. Let us notice that
$H_C$ is  {\em bilinear}
in $\sgr_i, \sgr_{i'}$. Then, the form: $
H_C=\sum_{i\in L,i'\in R}\ga_{i,i'} \sgr_i \cdot \sgr_{i'}
$ can be diagonalized by suitable linear transformation in the spin variables 
on "L" and the same transformation on "R". After diagonalization, $H_C$ 
takes the form 
\[
H_C = \sum_{I\in L, I'\in R} J_{I,I'} \sigr_I \cdot \sigr_{I'}
\]
where the matrix $ J_{I,I'}$ is diagonal and has only non-negative elements.
Then, we have system in the form analogous as ( 
\ref{PostacHC}) and one can repeat all considerations from previous
Subsection, obtaining analogous results. One should only remember that
the  "ice rule" concern spin variables after diagonalizing
transformation, i.e.  we have
\[
\langle \psi_0 | \sigr_I + \sigr_{I'} | \psi_0 \rangle=0.
\]


\section{Trace inequality and comparison of ground-state energies for various
  systems}
\label{SecCompar}
Lieb and Schupp have proved beautiful inequality for traces of matrices. It
is crucial in further considerations.

{\bf Theorem} (\cite{LS2}, \cite{S3}). For any square matrices $c,M,N$
it is true that
\be
|\Tr c^\dg M c N^\dg| \leq \frac{1}{2} (\Tr c_L M c_L M^\dg + \Tr c_R
N c_R N^\dg),
\label{MatrIneq} 
\ee
where $c_R=\sqrt{c^\dg c}$, $c_L=\sqrt{cc^\dg}$ are unique (positive)
square roots from positive definite matrices $c^\dg c$
and $cc^\dg$.

For the convenience of reader, the proof is supplied in the Appendix.
Now let us notice that the inequality (\ref{MatrIneq}) 
still remains valid if matrices
$c,M,N$ are {\em rectangular} ones: If $c$ is $m \times n$ matrix,
then  $M$ is  $m\times m$ matrix,  $N$ is 
 $n\times n$ matrix, $u$ is {\em partial} isometry, whereas $c_L$ and $c_R$
are positive matrices of dimension $m\times m$ and $n\times n$, respectively.

Now, let us consider the system {\em without} reflection symmetry,
which however still consists of the "left" subsystem $L$ described by
$H_L$ and the "right" subsystem $R$ described by $H_R$. 
The full system will be denoted by $L-R$, and its Hamiltonian
$H_{L-R}$  contains
also the "crossing bond" term  $H_C$, describing "interaction" of these two
parts: $H_{L-R} = H_L+H_R+H_C$. Every state of the system can be
written  in  the
form (\ref{ReprOfState}). We change slightly the notation and write an analogon
of  (\ref{ReprOfState}) as:
\be
\psi=\sum_{\alpha,\beta'}c_{\alpha \beta'} 
\psi^L_\alpha\otimes(\psi^R_{\beta'})_{\rm rot},
\label{ReprOfStateNonSym}
\ee
to distinguish between indices referring to the $L$ part (unprimed)
and $R$ (primed ones).
Remember that the $c$ matrix is now, in general,
the {\em rectangular} one.

Consider first the situation, where
$H_C$ contains only {\em one} bond:
\[
H_C= \sgr\cdot \Sgr,
\]
where $\sgr$ belongs to $L$ and $\Sgr$ belongs to $R$. 
One can write expression for the mean
value of energy in the manner analogous as Eq. (\ref{psiHpsi}):
\be
E_{L-R} = \Tr c c^\dg h_L + \Tr c^\dg c h_R 
- \sum_{\mu=1}^3 \Tr \left[ c^\dg t^{(\mu)} c (T^{(\mu)})^\dg\right],
\label{ELR}
\ee
where: $h_L$, $h_R$ are matrices of corresponding hamiltonians.
$t$ refers to the left subsystem, whereas $T$ to the right one; 
$t$, $T$ matrices are given by:
 $t^{(1,3)}_{\al\bet} = \langle \psi^L_\al |
s^{(1,3)}|\psi^L_\bet\rangle$,
$t_{\alpha\beta}^{(2)} = 
i \langle \psi_\alpha^L | s^{(2)}  |
\psi_\beta^L\rangle$, 
and similarly for $T$:
 $T^{(1,3)}_{\al'\bet'} = \langle \psi^R_{\al'} |
S^{(1,3)}|\psi^R_{\bet'}\rangle$,
$T_{\alpha'\beta'}^{(2)} = 
i \langle \psi_{\alpha'}^R | S^{(2)}  |
\psi_{\beta'}^R\rangle$. 

Inequality (\ref{MatrIneq}) applied to the last term of (\ref{ELR}) gives: 
\[
-\sum_{\mu=1}^3 \Tr c^\dg t^{(\mu)} c (T^{(\mu)})^\dg \geq 
- \frac{1}{2}\sum_{\mu=1}^3 
(\Tr c_L t^{(\mu)} c_L (t^{(\mu)})^\dg + \Tr c_R T^{(\mu)} 
 c_R (T^{(\mu)} )^\dg)
\]
Then, we can write:
\[
E_{L-R} \geq \Tr cc^\dg h_L + \Tr c^\dg c h_R - 
\frac{1}{2}\sum_{\mu=1}^3 (\Tr c_L t ^{(\mu)} c_L (t^{(\mu)} )^\dg 
+ \Tr c_R T^{(\mu)}  c_R (T^{(\mu)} )^\dg)
\]
\[
= \Tr c_Lc_L h_L + \Tr c_R c_R h_R - 
\frac{1}{2}\sum_{\mu=1}^3 (\Tr c_L t^{(\mu)}  c_L (t^{(\mu)}) ^\dg 
+ \Tr c_R T^{(\mu)}  c_R (T^{(\mu)} )^\dg)
\]
\be
= \frac{1}{2} (\Tr c_L c^\dg_L h_L +  \Tr c^\dg_L c_L h_L 
- \sum_{\mu=1}^3 \Tr c^\dg_L t^{(\mu)}  c_L (t^{(\mu)} )^\dg )
\label{almostE_L-L}
\ee
\be
+\frac{1}{2} (\Tr c_R c^\dg_R h_R + \Tr c^\dg_R c_R h_R
 - \sum_{\mu=1}^3 \Tr c^\dg_R T^{(\mu)}  c_R (T^{(\mu)} )^\dg ).
\label{almost E_R-R}
\ee
How can we interpret two last expressions?
They resemble very much ground-state energies for the following
systems: the  first one consists of two copies of $L$ subsystem 
 with "interaction"
Hamiltonian $H_C=\sgr\cdot\sgr'$,where $\sgr$ belongs to $L$, whereas
$\sgr'$  to its twin copy (let's denote it as the $L-L$ system). 
 The second one consists of two
copies of $R$ subsystem with  $H_C=\Sgr\cdot\Sgr'$ (it will be denoted
as  $R-R$ system).
 More precisely, they are energies of trial functions, built up
from matrices 
 $c_L$ and $c_R$, respectively. From variational principle, they are greater
than true ground-state energies, so we have the general inequality for
ground-state energies for three systems $L-R,L-L,R-R$:
\be
2 E_{L-R} \geq E_{L-L} + E_{R-R}. 
\label{2ELRvsELL+ERR}
\ee
Considerations above concerned the situation, where "interaction" part
was the only "bond" $\sgr\cdot\Sgr$. Generalization to multi-bond case is 
immediate. Let's have general "interaction" hamiltonian:
\[
H_C=\sum_{i\in L} \sum_{j'\in R} J_{ij'} \sgr_i \cdot \Sgr_{j'};
\]
where we assume that $J_{ij'}$ are positive numbers.
Then, the inequality (\ref{2ELRvsELL+ERR}) is still true for systems $L-L$
and $R-R$ with "interaction" hamiltonians: 
$ H_C = \sum_{i\in L} 
\left(\sum_{j'\in R} J_{ij'} \right) \sgr_{i}\cdot \sgr'_{i}$ for the
$L-L$  system, and
$ H_C = \sum_{j'\in R} 
\left(\sum_{i\in L} J_{ij'} \right) \Sgr_{j'}\cdot \Sgr'_{j'}$ for
$R-R$  system.

 Lieb and Schupp have applied inequality (\ref{2ELRvsELL+ERR}) 
to spin rings, getting
the following relation between ground-state energies $E_k$ for rings
with $k$  spins: 
\[
 2
  E_{n+m} \geq E_{2n} + E_{2m}
\]
 However, it seems that 
it can be used in much more general situations, if we consider the multi-bond
case.
\begin{figure}
\includegraphics{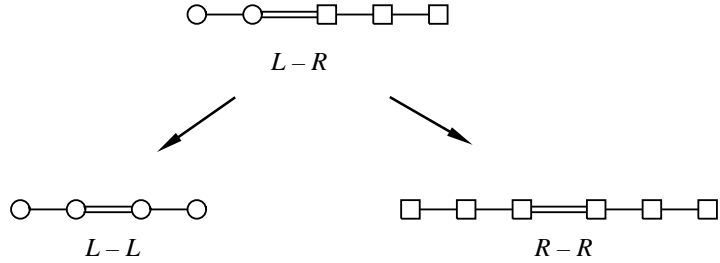}
\caption{\label{fig:fig4}  Illustration of inequality
  (\ref{Emn}) in the case of  spin chains for $n=2,m=3$.}
\label{fig4}
\end{figure}

{\bf Examples.}
\begin{enumerate}
\item For spin chains (Fig~\ref{fig4}.), we have analogous inequality:
\be
2 E_{n+m} \geq E_{2n} + E_{2m}.
\label{Emn}
\ee
\item Rather obvious is generalization of this result for systems
  defined on subsets of $\Zgr^d$ for $d>1$ (rectangles,
  parallelepipeds etc)  ($d=2$ case is
illustrated on FIG.~\ref{fig5}); as a result,
we again have inequality (\ref{Emn}), where $m,n$ are lengths of
systems in direction
perpendicular to reflection line (plane for $d=3$).
\begin{figure}
\includegraphics{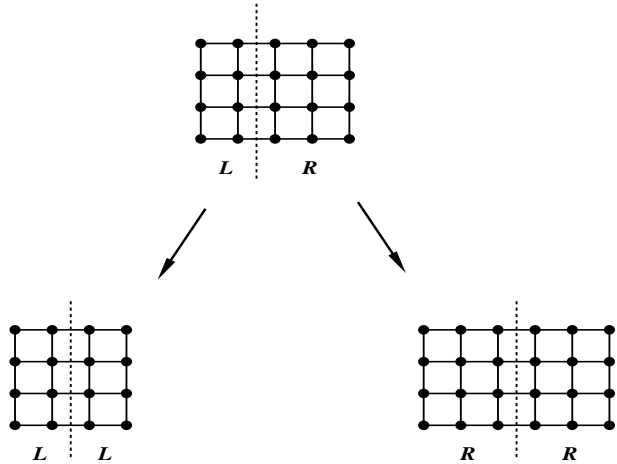}
\caption{\label{fig:fig5}  Illustration of inequality
  (\ref{Emn}) for 2d lattices; $m=2,n=3$.}
\label{fig5}
\end{figure}
\item Consider system defined on a rectangular subset of square
  lattice   (see FIG.~\ref{fig6}). Divide it into $L$ and $R$ part in
  nonsymmetric manner (in this example, this division is realized
  by the ``snaky'' line). Then, we obtain the inequality
  (\ref{2ELRvsELL+ERR}) for systems as pictured on  FIG.~\ref{fig6}. 
\begin{figure}
\includegraphics{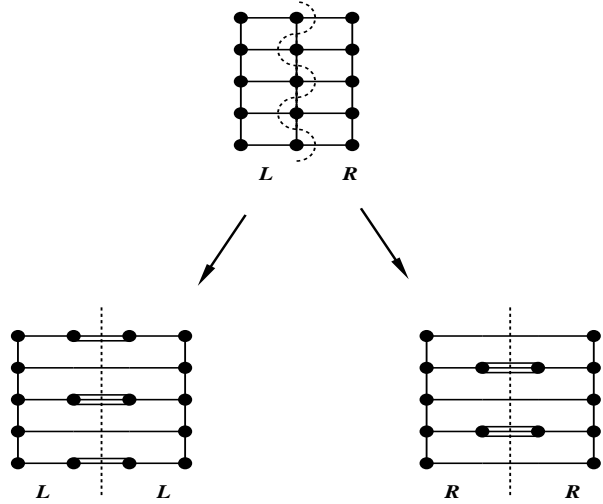}
\caption{\label{fig:fig6} Division of the system defined on a rectangular
  subset  of square lattice into $L$ and $R$ subsets by ``snaky''
  line. Single lines  denote coupling
  constants equal to $J$, double lines correspond to $2J$ couplings,
  and triple ones - to $3J$. The same convention is used on two
  following pictures.}
\label{fig6}
\end{figure}
\item Consider ``zig-zag ladder'' (see FIG.~\ref{fig7}). Divide it
  into parts $L$ and $R$ by the line going through the middle of the
  ladder and parallel to it. Then, the inequality
  (\ref{2ELRvsELL+ERR}) gives relation between ground-state energies
  of the ``zig-zag'' ladder and ordinary one, with suitable relation
  between coupling constants (see FIG.~\ref{fig7}).
\begin{figure}
\includegraphics{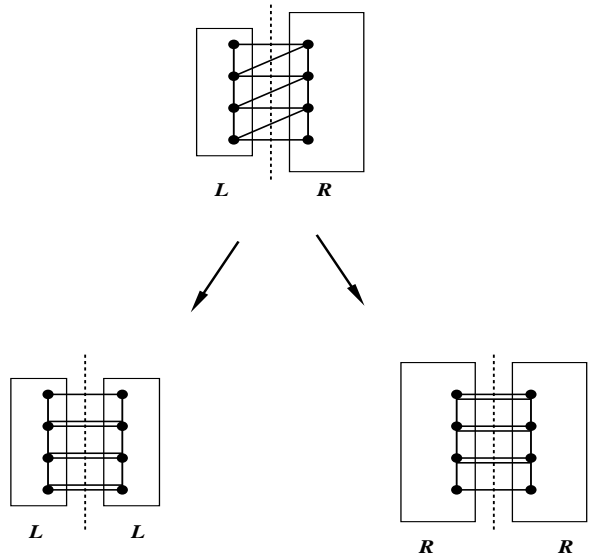}
\caption{\label{fig:fig7} ``Zig-zag'' and ordinary ladders.}
\label{fig7}
\end{figure}
\item The same construction can be applied to the ``pyrochlore''
  ladder (see FIG.~\ref{fig8}).
\begin{figure}
\includegraphics{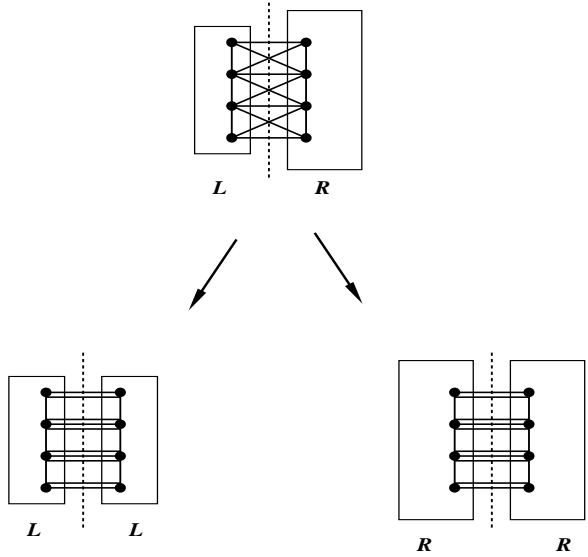}
\caption{\label{fig:fig8} ``Pyrochlore'' and ordinary ladders.}
\label{fig8}
\end{figure}
\item Ground-state energy relations for systems described in Example 4
  can be repeatedly used to obtain the inequality between systems
  defined on triangular and rectangular lattices. Namely, consider the
  system defined on isotropic triangular lattice with coupling
  constant $J$ and with periodic boundary
  condition in both (``horizontal'' and ``vertical'')
  directions. Assume,  moreover, that
  in the horizontal direction we have $2^k$ sites and $M$ in the
  horizontal one ($k,M$ are arbitrary). Let division of the
  system onto $L$ and $R$ parts will be realized by ``vertical'' plane
  in such a way that both parts contain equal number of sites.Then, by
  subsequent  use of inequality   (\ref{2ELRvsELL+ERR}), we obtain
  relation between ground state energies of systems on triangular
  lattice   $E_{\rm tr}$  and  on rectangular  one $E_{\rm rect}$:
\[
E_{\rm tr} \geq E_{\rm rect}
\]
where system on rectangular lattice has coupling constants equal $J$
in the vertical direction and $2J$ in the horizontal one.
\item Analogous construction can be applied to obtain the ground-state
  energy inequality
  between  systems  defined on lattices: isotropic pyrochlore one
  ($E_{\rm pyro}$)  and  rectangular  one ($E_{\rm rect}$):
\[
E_{\rm pyro} \geq E_{\rm rect}
\]
where  system on pyrochlore lattice has coupling constants all equal
to $J$, whereas those on rectangular lattice has couplings 
$J$ in the vertical direction and $3J$ in the horizontal one.
\end{enumerate}

\section{Summary}
\label{SecSumm}
In the course of paper, some extension and applications of Lieb-Schupp
approach have been given. The first result concerns relaxing of demand
of reflection symmetry, present in their papers. It turned out that
also other kinds of symmetry are allowed (inversion, $C_2$
rotation). The second result is an application of Lieb-Schupp
inequality, relating ground-state energies of various systems, to
numerous new (to my best knowledge) situations. The relation
between ground-state energies of models on triangular and rectangular
lattices can serve as an example. 

In the paper \cite{S3} Schupp wrote: ``There is no doubt that the
scheme can be further generalized''. I consider my paper as a step in
this direction, but of course possibilities of the Lieb-Schupp scheme
seem to be far from exhaustion. 
\appendix
\section{Appendix}
Here we supply proof of the inequality (\ref{MatrIneq}).

{\em Proof}: Using polar decomposition theorem, we can express the $u$ matrix
as:
 $c=uc_R$, where the $u$ matrix is unitary. We have:
$(uc_Ru^\dg)^2 = u c^\dg c u^\dg = (uc_R) (c_Ru^\dg) =
(uc_R)(uc_R)^\dg= cc^\dg = c_L^2$, then, because the (positive) square
root is unique, we have:
$uc_Ru^\dg=c_L$. Analogously, for arbitrary analytic function $f$ defined
on positive real numbers, we have:
$uf(c_R)u^\dg = f(c_L)$. In particular, $u\sqrt{c_R}=\sqrt{c_L}u$,
which implies: $c=\sqrt{c_L}u\sqrt{c_R}$. Now, let: $P= u^\dg
\sqrt{c_L} M \sqrt{c_L}u$ and $Q:=\sqrt{c_R}N^\dg\sqrt{c_R}$; then, we have: 
\[
|\Tr c^\dg M c N^\dg| = |\Tr PQ| \leq \frac{1}{2} (\Tr  PP^\dg + \Tr  QQ^\dg)
\]
\be
 =\frac{1}{2} (\Tr  c_L M c_L M^\dg + \Tr  c_R N c_R N^\dg),
\ee
where we have used Schwarz inequality for matrices:
\[
|\Tr  PQ| = |\sum_{i,j}P_{ij}Q_{ji}|
\leq \frac{1}{2}\sum_{i,j}(|P_{ij}|^2 + |Q_{ji}|^2)
=\frac{1}{2}(\Tr PP^\dg + \Tr QQ^\dg).
\]
\noindent {\bf Acknowledgements.}
 This work was supported by  the 
Polish Research Committee (KBN) under  Grants: No.~2~P03B~131~19 and
110\slash 501\slash SPUB\slash 27, and
by the Postdoctoral Training Program HPRN-CT-2002-00277.

\end{document}